\documentstyle[12pt,aaspp4,epsfig]{article}

\newcommand{\be}{\begin{equation}}
\newcommand{\ee}{\end{equation}}
\newcommand{\lab}[1]{\label{#1}}
\newcommand{\rr}[1]{~(\ref{#1})}
\newlength{\oneofour}
\newcommand{\veconeofour}{\mbox{\raisebox{\oneofour}{$\longrightarrow$}
$\!\!\!\!\!\!\!\!\!\!\! 104$}}

\begin{document}

\title{Low Microlensing Optical Depth Toward the Galactic Bar}

\author{Piotr Popowski}
\affil{MPA, Karl-Schwarzschild-Str. 1, Postfach 1317, 85741 Garching
bei M\"{u}nchen, Germany}
\affil{E-mail: popowski@mpa-garching.mpg.de}

\begin{abstract}
I make a new evaluation of the microlensing optical depth toward the 
Galactic bar
from Difference Image Analysis (DIA) of the MACHO Collaboration. 
First, I present supplementary evidence that MACHO field 104 located at 
$(l,b) = (3\hbox{$.\!\!^\circ$}11,-3\hbox{$.\!\!^\circ$}01)$ is
anomalous in terms of the event duration distribution.
I argue that both the event durations and the very high optical depth 
of field 104
are not representative and, therefore, exclude this field as an
outlier.
In addition, I eliminate field 159 at $(l,b) =
(6\hbox{$.\!\!^\circ$}35,-4\hbox{$.\!\!^\circ$}40)$ based mainly
on its separate location, but also on unexplained statistical
properties of the event durations.
The remaining six DIA fields form a very homogeneous and spatially
compact set that is very suitable for averaging.  
The weighting of the optical depth values for these six DIA
fields results in a total optical depth $\tau_{tot} = 2.01
^{+0.34}_{-0.32} \times
10^{-6}$ at $(l,b) = (2\hbox{$.\!\!^\circ$}22,-3\hbox{$.\!\!^\circ$}18)$.
If a fraction of all sources, $f_{\rm disk}$, assumed to be in the
disk, does not contribute to 
microlensing, then the optical depth toward the sources in the bar is
$\tau_{\rm bar} = 2.23^{+0.38}_{-0.35} \times
10^{-6} \, \frac{0.9}{1-f_{\rm disk}}$.
Both $\tau_{\rm tot}$ and $\tau_{\rm bar}$ are substantially
lower than the original estimates of Alcock et al.
Most of the change in the DIA-based optical depths comes
from a more appropriate statistical treatment of the results
in individual fields and not from the removal of fields 104 and 159.
When taken together with $\tau_{\rm bar} = 1.4 \pm 0.3 \times
10^{-6}$ at 
$(l,b) = (3\hbox{$.\!\!^\circ$}9,-3\hbox{$.\!\!^\circ$}8)$ as derived 
from clump
giants, this new result suggests that the conclusions from
microlensing experiments are in reasonable agreement with expectations from 
infrared-based Galactic models.

Subject Headings: Galaxy: center --- Galaxy: structure --- gravitational lensing
\end{abstract}

\section{Introduction}

The structure and composition of our Galaxy is one of the 
outstanding problems in contemporary astrophysics. 
Microlensing is a powerful tool
to learn about massive objects in the Galaxy.
The amount of matter between the source and observer
is customarily described in terms
of the microlensing optical depth, which is defined as the probability that a 
source flux will be gravitationally magnified by more than a factor of
1.34.
The optical depth, $\tau$, is typically estimated using the following formula:
\be
\tau = \frac{\pi}{2NT} \sum_{\small\rm all \; events} \frac{t_E}{\epsilon(t_E)}, \lab{optdepth}
\ee
where $N$ is the number of observed stars (potential sources), $T$ is
the total exposure (roughly equal to the temporal span of
observations), and 
$\epsilon(t_E)$ is the efficiency for detecting an event with a given 
Einstein radius crossing time $t_E$.

Early microlensing analyses (Udalski et al.\ 1994; Alcock et al.\
1997) of the lines of sight toward the Galactic center produced two 
unexpected results: (1) a very high optical depth of $3-4 \times
10^{-6}$, exceeding expectations from Galactic models
by a factor of a few and (2) an 
overabundance of long-lasting events.
These two issues indicated that either we still lacked a basic understanding
of the Galactic structure and/or that the microlensing results needed revision.
The second conclusion was certainly true, since 
these old analyses suffered either from small number statistics or 
from oversimplified analyses of the detection efficiency of microlensing
events.

To describe the current situation, I will use the recent results from
the MACHO Collaboration. The main goal of the MACHO Project
was to discover and characterize dark matter and other faint objects
through detection and analysis of microlensing events seen toward 
the Magellanic Clouds and the central region of the Milky Way.
The MACHO observations were performed with the $1.27$-meter telescope at 
Mount Stromlo Observatory, Australia. 
A detailed description of the MACHO
telescope and photometry is given in Alcock et al.\ (2001).
In total, MACHO collected seven seasons (1993-1999) of two-filter data
in 94 Galactic bulge fields, which are of interest here.

Blending is a major concern in any analysis of the 
microlensing data. The bulge fields are crowded, so that
the objects observed at a certain atmospheric seeing are blends of several 
stars, of which only one is typically lensed.
This situation complicates the determination of an event's parameters
and the estimate of the detection efficiency of microlensing events.
There are two major approaches to minimize these problems:
(1) One may work with bright stars like clump giants that are subject
to little blending and can be utilized without the knowledge of the
stellar luminosity function down to faint magnitudes. Such an analysis was
performed by Popowski et al.\ (2000, 2001), who calculated the optical depth
based on data from five seasons (1993-1997) in 77 fields.
(2) One may improve the photometry and minimize blending by analyzing
only the varying part of flux. This is accomplished by subtracting
images of the same fields taken on different nights.
This strategy was realized by Alcock et al.\
(2000, hereafter AD2000), who analyzed three seasons (1995-1997) of data in
8 MACHO fields.

AD2000 determined a new value of the optical depth based on the
Difference Image Analysis (DIA) technique that included a detailed 
computation of the detection efficiencies. They estimated
the optical depth toward the Galactic bar, $\tau_{\rm bar} =
3.23^{+0.52}_{-0.50} \times 10^{-6}$ at  
$(l,b) = (2\hbox{$.\!\!^\circ$}68,-3\hbox{$.\!\!^\circ$}35)$.
This value was declared to be too high to be reconciled with
the constraints coming from the Galactic rotation curve 
and the local density of stars (Binney, Bissants, \& Gerhard 2000;
Bissantz \& Gerhard 2002). To shed more light on this problem, I will
reconsider DIA data using the characteristics
of 99 DIA events from AD2000 reported in their Tables 4, 5, and 6.

The purpose of this paper is to critically discuss current
observational constraints on the microlensing optical depth toward
the Galactic bar. I suggest slight modifications of the analyzed samples
and appropriate methods of data averaging. 
First, in \S 2, I point to some unusual properties of one
of the MACHO fields. I show that, despite completely different
analysis methods and different event selection criteria, MACHO field 104
is a clear outlier in both clump giants analysis and DIA (Popowski et
al.\ 2001; AD2000). The exclusion of this field does not change DIA
results significantly, but has far-reaching consequences for the
optical depth based on clump giants.
I argue in \S 3 that excluding field 159 from DIA will result in a more
spatially compact group of fields better suited for averaging.
In \S 4, I consider statistical properties of errors in optical depth of
individual fields. I show in \S 5 that the weighted average of all
individual DIA fields (with no exclusions!) produces a 20\% lower optical
depth than the AD2000 value.
When fields 104 and 159 are excluded, the optical depth from DIA remains
roughly the same but is given at a location closer to the Galactic center.
I also discuss the transition from the total observed optical depth,
$\tau_{\rm tot}$, to the one toward sources residing in the Galactic
bar, $\tau_{\rm bar}$. I conclude in \S 6 comparing the adjusted
optical depths from clump giants and DIA with the recent results
from Evans \& Belokurov (2002) and Bissantz \& Gerhard (2002).

\section{Field 104}

Popowski et al.\ (2000, 2001) argued that MACHO field 104 located at
$(l,b) = (3\hbox{$.\!\!^\circ$}11,-3\hbox{$.\!\!^\circ$}00)$ is very unusual from the microlensing
perspective when analyzed using events with clump giants as sources.
They noticed that one fifth of the $\sim$ 50 MACHO
clump events are in field 104,
and that there is a high concentration of long-duration events in this
field (5 out of the 10 events longer than 50 days are in 104, including
the longest 2).

I will now review part of Popowski's et al.\ (2000, 2001) work to present
one of the possible methods for the analysis of duration
distributions.
One can compare the distribution of event durations in field 104 and
all the other fields.
Ideally, one would like to account for the change
in the efficiency of event detection with different durations in
different fields. However, using only uncorrected
event durations one can still place a useful, lower limit on
the significance of this difference. The efficiency for detecting long
events should be similar in most fields, because it does not depend strongly 
on the sampling pattern. In contrast, the detection efficiency for
short events will be lower in sparsely sampled fields. Therefore,
the number of short events in some of the fields used for comparison
(the
ones that are sparsely sampled) may be
relatively too small with respect to the frequently-sampled field 104.
Proper accounting for this efficiency difference, however, would only 
increase the
significance of the $t_E$ distribution difference.
In conclusion, the analysis of event durations {\em uncorrected} for 
efficiencies should provide a lower limit on the difference between field 
104 and all the remaining fields.

The Wilcoxon's number-of-element-inversions statistic is well suited to test
whether events in 104 and other fields can originate from the same
population.
First, one separates the events into two samples: events in 
field 104 and all
the remaining ones.
Second, one orders the events in the combined sample from the 
shortest to the longest. Then, being only allowed to swap the adjacent
events, one counts
how many times one would have to exchange the events from field 104 with
the others to have all the 104 field events at the beginning of the list. 
If $N_1$ and $N_2$ designate numbers
of elements in the first and second sample, respectively, then for $
N_1 \geq 4$, $N_2 
\geq 4$, and $(N_1+N_2) \geq 20$, the Wilcoxon's statistic is approximately
Gaussian distributed with an average of $N_1 N_2/2$ and a dispersion $\sigma$
of $\sqrt{N_1 \, N_2 \, (N_1+N_2+1) / 12}$.
Popowski et al.\ (2000) found for the clump sample that the events in
104 differ (are longer) by $2.55 \sigma$ from the other fields.

Here I argue that the analysis of the DIA events taken from AD2000, which
are 
of general type, confirms
the clump-based conclusions.
I note, however, that a strict mathematical interpretation of this
result requires some caution as the clump and DIA samples
share 7 events, 4 of which are in field 104.
The Wilcoxon's statistic for the DIA sample, split into field 104 and the
rest, is equal to 904,
whereas the expected number for subsamples drawn from the same
parent population is 648 with an error of about 103. Therefore,
the events in 104 differ (are longer) by $2.49 \sigma$ from the other
fields.

Figure 1 presents a comparison between event durations in field 104 and 
the rest of the 
DIA sample using 97 out of 99 events. The remaining 2 events are
not included as they were classified by AD2000 as binaries.
The upper panel shows the number of events as a function of event
duration based on all 8 DIA fields.
Field 104, represented by the black portion of the histogram, clearly
dominates the long duration part of this distribution.
The middle panel presents a comparison of the number distribution
functions (normalized)
in field 104 and the remaining seven fields. The distribution in field
104 is much flatter and more extended. The average durations, $<t_E>$, for both
distributions are marked with vertical arrows. In field 104,
$<t_E> = 48 \pm 10$, in the remaining fields $<t_E> = 24 \pm 2$.
The lower panel is a plot of cumulative distributions of durations
for field 104 and the remaining fields. Kolmogorov-Smirnov test
indicates that the probability that both samples come from the same
parent population is $P = 0.0335$.

There is no doubt that the
significance of this effect depends on selection criteria of events.
However, the fact that the effect is so significant in two weakly
overlapping and differently processed samples, makes it more reliable.
In addition, the unique character of field 104 is supported 
by its optical depth.
The optical depth is 10 times higher than
the average of the other fields as estimated from clump giants (more
than $2 \sigma$ effect) and
2 times higher based on DIA (low significance).
Galactic models typically do not account for very localized structures,
so they have no way to explain fields like 104. When included, structure
like this will bias the results toward a different model.

The above analysis reinforces Popowski's et al.\ (2001) suggestion to
treat field 104 as an outlier and exclude it from the determination 
of the optical depth. I conclude that in the case of clump giants,
$\tau_{\rm bar} = 1.4 \pm 0.3 \times 10^{-6}$ is preferred over
the value of $\tau_{\rm bar} = 2.0 \pm 0.4 \times 10^{-6}$ based on
the entire sample in 77 fields.

When field 104 is removed the position at which the clump giant optical
depth is evaluated does not change significantly. One may see this 
from the following considerations. Let $\vec{x}_{\rm new}$ indicate
the new position, $\vec{x}_{\rm old}$ the old average position for all 
77 clump fields, and \veconeofour $\;$ the position of field 104.
Then we have
\be
\vec{x}_{\rm new} = \frac{\vec{x}_{\rm old} - f \veconeofour}{1-f}, \lab{newpos}
\ee
where $f$ is the weight of field 104. 
The exact value of $f$ depends on the weighting scheme, but $f$ is
going to be small as one can realize looking at 
two representative choices for
$f$: (1) $f = 1/N_{\rm fields} = \frac{1}{77} \approx 0.013$, as 77
fields were analyzed by Popowski et al.\ (2001) (2) $f =
N_{*,104}/N_{*, {\rm all \; fields}} \approx 0.028$, where $N_{*,\cdot}$
indicates the number of clump giants in the region described by the second
subscript i.e., ``104'' or ``all fields''. In both cases, 
the position at which the new optical depth
is determined does not change within the accuracy of the original
value, $(l,b) = (3\hbox{$.\!\!^\circ$}9, -3\hbox{$.\!\!^\circ$}8)$.

\section{Field 159}

Figure 2 displays the spatial distribution of eight fields used by AD2000
(analog of their Figure 5).
Field 159 centered at $(l,b) =
(6\hbox{$.\!\!^\circ$}35,-4\hbox{$.\!\!^\circ$}40)$ is clearly separated from the others. When averaged together
with the other fields it will substantially influence the effective position
at which optical depth is determined.
The filled circle indicates the average weighted position\footnote{The most
appropriate way to compute an average position is discussed in \S 5.} of 6
fields, excluding fields 104 and 159. The average weighted
position of 8 fields is marked with a cross for comparison. The open square
indicates the average position for optical depth of $\tau_{\rm bar} =
1.4 \pm 0.3$ based on clump giants (Popowski et al.\ 2001).
Based on its separate location, and somewhat unusual properties
of the errors in the optical depth (see next section), I conservatively 
remove field 159 from the further analysis.

\section{Statistical considerations}

Table 1 provides additional support of my previous suggestion to
remove fields 104 and 159 from the final determination.
Columns 1--3 are taken directly from AD2000 and list a
field, the number of events in each field, and the corresponding
DIA-based optical depth, respectively.
Han \& Gould (1995) showed that in general the error in an optical depth
can be split into two terms: one which is the simple Poisson noise in
the number of events and another one that comes from the dispersion in the
efficiency corrected duration distribution. Han \& Gould (1995) used
an example of constant efficiency to argue that this second term
can be quite substantial. In reality, the shape of the intrinsic duration
distribution and the shape of the detection efficiency curve are quite
similar and conspire to produce a rather low dispersion when divided
by each other. Therefore, one expects that the ratio between the 
total error in $\tau$ and the error 
based only on Poisson statistic will be larger than unity but not by
much.
When I examined errors in the optical depth reported in Table 3 of
AD2000, I realized that several of them were of the order or even
somewhat smaller than one would expect based on the Poisson distribution
alone.
Indeed, it turned out that some of the errors needed small adjustments,
and the {\em correct} errors (Andrew Drake; private
communication) are reported in columns 4 and 5 of Table 1.
Column 6 gives a Gaussian approximation to 1-sigma Poisson errors 
in the optical depth.
Finally, columns 7 and 8 list the ratios of the true error obtained
with bootstrapping technique by AD2000 to that expected from 
the Poisson distribution alone.

I concentrate on the last two columns of Table 1.
First, let me notice that now all positive and negative errors have larger
absolute values than their Poisson counterparts, which is
expected based on Han \& Gould (1995) argument.
Second, the examination of individual values for different fields
indicates that both fields 104 and 159 are different from the rest.
This different character is particularly well established in the case
of fields 104 with 16 events, and somewhat less obvious for field 159
with only 4 events, where the ratios in columns 7 and 8 may be
slightly affected by my Gaussian approximation to Poisson confidence
intervals.
The large values of the error ratios in field 104
suggest that dispersion in $t_E/\epsilon(t_E)$ contributes significantly to
the errors in optical depth in this field, implying an
unusual duration distribution.
In summary, the decision to exclude fields 104 and 159 gains additional
justification.

\section{Optical depth}

The microlensing optical depth can be used to distinguish between 
different Galactic models. Most models share the same
major traits, and so the optical depth must be both accurate and
precise to render a definitive answer.
Precision depends mostly on the number of detected events and,
therefore, is predetermined for a sample of a given size.
Accuracy of the result depends mostly on two factors: systematic
errors and methods of analysis.

Here I would like to describe one bias that is very hard to eliminate
when one obtains a single value of the optical depth 
based on a ``gradient-large'' region. ``Gradient-large'' means 
that the gradient of the optical depth changes
significantly within this region.
Suppose that one observes a number of small fields in this region 
and determines
their optical depths $\tau_i \pm \sigma(\tau_i)$. Then, the most
natural way to obtain a model-independent, single value
of the optical depth representative
for the region will be a weighted average of the results in individual
fields (see also the discussion below). However, this method implicitly
assumes that the optical depth changes in a linear fashion.
This is inconsistent with the fact that the gradient in the optical depth
changes across the region.
The MACHO fields in the Galactic bar that are spread in an area of a
few by a few degrees constitute a gradient-large region
as can be seen e.g., from microlensing maps by Evans \& Belokurov
(2002). It is worth noting, however, that in this region
some models are much more linear than the others (e.g., Freudenreich's 1998
versus Binney, Gerhard, \& Spergel's 1997). Nevertheless, in most
cases this type
of simple averaging will produce a bias. Therefore, the best solution is
to directly compare the models with values in individual fields
using the maximum likelihood method. Still, averages may be used
for illustrative purposes and as approximate solutions in the case when
field-by-field detection efficiencies are not accessible.

In addition,
even in the absence of the above non-linearity bias, the
average
optical depth based on global efficiencies in all fields taken together
is also biased. On average there are more events per source star in fields with
higher optical depth. On the other hand, an average position at which
the detection efficiency is evaluated depends only on the number of
sources in different fields.
These two weightings produce a mismatch between the average position of
events and the average position for universal efficiency. This type 
of bias might have affected the clump giants determination, but here I am
not able to improve upon the original Popowski et al.\ (2001) results as 
the final field-by-field data for clump giants are not available yet.

Let us now turn to the study by AD2000.
First, one should notice that the optical depth reported by AD2000,
even though based on microlensing efficiencies in eight individual
fields, is not equal to the average of values given in their Table 3. 
The average optical depth from AD2000 was computed based on a ``hybrid'' type
of prescription. The procedure was global in the sense that 
equation.\rr{optdepth} for the optical depth was used for the entire sample of events
and $N$ was taken to be the number of all relevant stars observed
in the eight fields. However, the procedure was also local in the sense
that efficiencies correcting the durations were taken from individual
fields.
An average of the optical depths in individual fields determined in a 
{\em consistent} fashion yields a less biased solution.
I compute a weighted average based on all eight fields and obtain:
(1) $\tau_{\rm 8 \; fields} = 1.98 \times 10^{-6}$ with $\sigma_{+}
=0.26 \times 10^{-6}$ using 
positive errors only\footnote{In what follows, I will refer to the error
that would increase the value of $\tau$ as to the positive one, and to
the error that would decrease the value of $\tau$ as to the negative one.}
and (2) $\tau_{\rm 8 \; fields} = 
1.96 \times 10^{-6}$ with $\sigma_{-} = 0.23 \times 10^{-6}$ using
negative errors only.
This can be summarized as $\tau_{\rm 8 \; fields} =
1.97^{+0.26}_{-0.23} \times
10^{-6}$, where the average position of 8 fields is the
same as in AD2000: $(l,b) = (2\hbox{$.\!\!^\circ$}68,-3\hbox{$.\!\!^\circ$}35)$. The ``weighted'' optical
depth $\tau_{\rm 8 \; fields}$ is nearly 20\% smaller than the value $\tau =
2.43^{+0.39}_{-0.38} \times 10^{-6}$ given in AD2000.

The important problem one faces is how to estimate the appropriate
position at which the optical depth is given when computed as an average
of values in individual fields. There are three major possibilities:
\begin{enumerate}
\item an average of central positions of the fields with no weighting 
(e.g., AD2000),
\item a weighted average of central positions with the optical depth errors
as weights,
\item a weighted average of central positions with the number of stars in
different fields as weights.
\end{enumerate}
Let me consider two limiting cases that will show the preferred
solution. In both cases, I consider $N$ fields with optical depth
measurements, $\tau_i$.
First, let the error in $k$-th field, $\sigma(\tau_k)$ be small, and let
$\sigma(\tau_i)/\sigma(\tau_k) \longrightarrow \infty$ for $i \neq k$.
It is obvious that one has no information about the optical depth at
the unweighted average position of all fields. However, with
prescription 1. one will be forced to report $\tau_k$ at this average
location. Prescription 2. will correctly give $\tau_k$ at the position 
of the $k$-th field.
Second, let errors in all $\tau_i$ be identical, and let
$\tau_k = C_1$ and $\tau_i = C_2$ for $i \neq k$ with $C_1 \neq C_2$.
Moreover let $N_k/N_i \longrightarrow \infty$ for $i \neq k$, 
where $N_i$ designates the number of stars in $i$-th field.
Prescription 3. will then report a value $C_2 + \frac{C_1-C_2}{N}$ at 
the location of the $k$-th
field leading to a disagreement with optical depth $\tau_k = C_1$.
I conclude that prescription 2. is the most appropriate.
Accordingly, the optical depth for 8 DIA fields computed with the new,
corrected errors is given at
$(l,b) = (2\hbox{$.\!\!^\circ$}68,-3\hbox{$.\!\!^\circ$}31)$, which 
accidentally is almost identical to the original unweighted value.

If one conservatively removes fields 104 and 159, the results from the
remaining six fields are as follows:
1. $\tau_{\rm tot} = 2.01 \times 10^{-6}$ with $\sigma_{+}
=0.28 \times 10^{-6}$ using positive errors only,
2. $\tau_{\rm tot} = 2.01 \times 10^{-6}$ with $\sigma_{-}
=0.25 \times 10^{-6}$ using negative errors only.
This can be summarized as $\tau_{\rm tot} = 2.01^{+0.28}_{-0.25} \times
10^{-6}$,
which at the first glance looks almost identical to the previous
result. However,
effectively this is a lower optical depth, as the
average weighted position of the six fields is closer to the Galactic
center at $(l,b) = (2\hbox{$.\!\!^\circ$}22, -3\hbox{$.\!\!^\circ$}18)$ 
[average unweighted position is $(l,b) = (2\hbox{$.\!\!^\circ$}00,
-3\hbox{$.\!\!^\circ$}23)$].
If, following AD2000, I add in quadrature the $10\%$ uncertainty in the
luminosity function, then
\be
\tau_{\rm tot} = 2.01^{+0.34}_{-0.32} \times
10^{-6} \;\;\;\;\; {\rm at} \;\;\;\;\; 
(l,b) = (2\hbox{$.\!\!^\circ$}22, -3\hbox{$.\!\!^\circ$}18), \lab{tautotfinal}
\ee
Unlike for clump giants, in the case of DIA, the removal of field 104
has little effect on the optical depth! Almost the entire change in
the optical depth comes from replacing the ``hybrid'' procedure of
AD2000 with a proper weighting of values in individual fields.

The value $\tau_{\rm tot}$ is the total optical depth toward the bar where all
stars that belong to different populations (predominantly bar and
disk) can be both sources and lenses. Oftentimes, one is interested
in the optical depth toward sources that reside in the bar only,
$\tau_{\rm bar}$.
This quantity is the easiest to evaluate directly for a well defined group of
sources that are much more abundant in the bar than in the disk.
Clump giants are believed to be such a population.
The events detected by DIA do not share this property. Fortunately,
under certain assumptions, we can also evaluate $\tau_{\rm bar}$ indirectly
based on $\tau_{\rm tot}$.
Let me assume that:
(1) microlensing detection efficiency does not depend on the membership
   in a given stellar population,
(2) the contribution of events with a disk star as a source to $S \equiv
\sum_{\rm all \; events}
\frac{t_E}{\epsilon(t_E)}$ is negligible.
Then, assuming that the disk and the bar are the only relevant populations, one
has the following relations:
\be
\tau_{\rm tot} = \frac{\pi}{2 (N_{\rm bar}+N_{\rm disk}) T} S, \lab{tautot}
\ee
and
\be
\tau_{\rm bar} = \frac{\pi}{2 N_{\rm bar} T} S, \lab{taubar}
\ee
which imply that
\be
\tau_{\rm bar} = \frac{N_{\rm bar}+N_{\rm disk}}{N_{\rm bar}} \tau_{\rm tot} =
\frac{1}{1-f_{\rm disk}} \tau_{\rm tot}, \lab{taurelation}
\ee
where $f_{\rm disk} \equiv \frac{N_{\rm disk}}{N_{\rm bar}+N_{\rm
disk}}$. 
This is a classical correction applied by Alcock et al.\ (1997) and
AD2000. In this case,
$f_{\rm disk}$ does not describe a contribution of any population to the
optical depth\footnote{Note that Alcock et al.\ (1997) discussion of the 
relation between $\tau_{\rm bar}$ and  $\tau_{\rm tot}$ is clearer than 
the one in AD2000.}, but rather the effect of ``dilution'' of the optical
depth caused by the foreground sources. Alcock et al.\ (1997) took 
$f_{\rm disk} = 0.20$ adjusting infrared estimates from the
DIRBE maps of Weiland et al.\ (1994) to the optical.
Probably following Alcock et al.\ (1997), AD2000 took even somewhat 
higher value
of $f_{\rm disk} = 0.25$. However, from the definition of $f_{\rm
disk}$, one sees that it should be approximately equivalent to a factor $p$
that was estimated
by AD2000 for each field and is given in column 8 of their Table 3.
On average $p \approx 0.1$, with very little scatter
around this value. Therefore, in what follows, I will take 
$f_{\rm disk} = 0.1$ as more appropriate.
Consequently, equations\rr{tautotfinal} and\rr{taurelation} imply that:
\be
\tau_{\rm bar} = 2.23^{+0.38}_{-0.35} \times
10^{-6} \; \frac{0.9}{1-f_{\rm disk}}\;\;\;\;\; {\rm at} \;\;\;\;\; 
(l,b) = (2\hbox{$.\!\!^\circ$}22, -3\hbox{$.\!\!^\circ$}18), \lab{taubarfinal}
\ee
where, similarly to AD2000, I computed the final error adding in
quadrature 10\% error originating from uncertainty in the
luminosity function. The values of $\tau_{\rm bar}$ for individual
fields are listed in Table 2. The errors in optical depths in columns 6 and
7 do not include additional error in the luminosity function.
Boldfaced are the fields that have
been used in my final determination of the optical depth. 
 
\section{Conclusions}

I have critically assessed the current observational situation regarding 
the microlensing optical depth determinations toward the Galactic bar
using the results from the two recent analyses by the MACHO Collaboration
(AD2000; Popowski et al.\ 2000, 2001).
First, based on 97 events from DIA,
I confirm the unusual character of the MACHO field 104 located at
$(l,b) = (3\hbox{$.\!\!^\circ$}11,-3\hbox{$.\!\!^\circ$}00)$ previously documented by Popowski et al.\ (2001), and reinforce
their suggestion to disregard this field as an outlier.
As a result, clump giant analysis produces $\tau_{\rm bar} = 1.4 \pm 0.3 \times 10^{-6}$ at
$(l,b) = (3\hbox{$.\!\!^\circ$}8, -3\hbox{$.\!\!^\circ$}9)$ from 76
fields with field 104 eliminated. This value
seems to be better motivated than 
$\tau_{\rm bar} = 2.0 \pm 0.4 \times 10^{-6}$ from the entire clump
sample.

Second, I make three important adjustments to DIA by AD2000.
\begin{itemize}
\item I estimate the average optical depth by weighting the optical depths in
individual fields using the optical depth errors as weights. 
This replacement of AD2000's ``hybrid'' procedure
reduces the optical depth by almost 20\%.
\item I eliminate 2 out of 8 fields: 104 as an outlier in duration distribution and
optical depth and field 159 based on its separation from the other
DIA fields. I also discuss the preferred determination of the position
at which the optical depth is reported.
\item I argue for a smaller ``dilution'' factor, $f_{\rm disk}$,  that
allows one to
   convert between the total measured optical depth, $\tau_{\rm tot}$ and 
the optical depth toward sources in the bar, $\tau_{\rm bar}$. 
This correction alone reduces the optical depth by 17\%.
\end{itemize}
The final DIA-based results are
$\tau_{\rm tot} = 2.01^{+0.34}_{-0.32} \times
10^{-6}$ and
 $\tau_{\rm bar} = 2.23^{+0.38}_{-0.35} \times
10^{-6}$
for $f_{\rm disk} = 0.1$, both at 
$(l,b) = (2\hbox{$.\!\!^\circ$}22, -3\hbox{$.\!\!^\circ$}18)$.

These new estimates can be compared with the most recent modeling
results from Evans \& Belokurov (2002) and Bissantz \& Gerhard (2002).
Evans \& Belokurov (2002) considered three
Galactic models by Binney et al. (1997), Dwek et al.\
(1995), and Freudenreich (1998), and included
the effects of spiral structure and streaming motions in the Galactic
bar.
Bissantz \& Gerhard (2002) constructed a non-parametric model with
spiral arms.
Based on the previous higher
estimates of the optical depth, Evans and Belokurov (2002) suggested that
Freudenreich's (1998) model is preferred, although
at a rather low significance.
Table 3 presents a comparison of various Galactic models with 
the optical depths derived here. 
The optical depth values for the clump location are taken directly from
Evans \& Belokurov (2002) or Bissantz \& Gerhard (2002), whereas the 
unpublished values at the DIA
location were kindly provided by Wyn Evans and Ortwin Gerhard. 
The numbers in parentheses
include the effect of the spiral structure.
Note that such comparison is only qualitative due to:
(1) possible biases in average values of the optical depth, and
(2) inconsistency between the disk model indirectly implied by the AD2000
luminosity function and the disk model used by Evans \& Belokurov
(2002) or derived by Bissantz \& Gerhard (2002).
Nevertheless, Table 3 suggests that the new, lower values of
the microlensing optical depth derived here are consistent
with the infrared-based models of the Milky Way but 
do not clearly favor a specific Galactic model.

\acknowledgments

I thank Andrew Drake for illuminating discussions about the issues
considered in this paper and his comments to the original draft
of this manuscript. The detailed comments and suggestions
made by Greg Rudnick substantially improved the presentation of
the results. I am very grateful to Wyn Evans and Ortwin Gerhard
for providing unpublished values of the optical depth for different 
Galactic models.

\begin{deluxetable}{cccccccc}
\tablecaption{Errors in the Optical Depth for Individual Fields\label{table1}}
\tablewidth{0pt}
\tablehead{
\colhead{Field} & 
\colhead{$N_{DIA}$} &
\colhead{$\tau$ ($\times 10^{6}$)} &
\colhead{$\sigma_{+}$} &
\colhead{$\sigma_{-}$} &
\colhead{$\sigma_{G}$} &
\colhead{$\frac{\sigma_{+}}{\sigma_{G}}$} &
\colhead{$\frac{\sigma_{-}}{\sigma_{G}}$}
}
\startdata
101 & 11 & 1.72 & 0.61 & 0.54 & 0.52 & 1.18 & 1.03 \nl
104 & 16 & 4.18 & 1.65 & 1.37 & 1.04 & 1.57 & 1.31 \nl
108 & 16 & 2.39 & 0.68 & 0.61 & 0.60 & 1.13 & 1.03 \nl
113 & 17 & 1.96 & 0.54 & 0.49 & 0.47 & 1.14 & 1.03 \nl
118 & 13 & 2.64 & 0.97 & 0.84 & 0.73 & 1.33 & 1.15 \nl
119 & 12 & 2.43 & 0.95 & 0.81 & 0.70 & 1.35 & 1.15 \nl
128 & 10 & 1.62 & 0.62 & 0.54 & 0.51 & 1.21 & 1.06 \nl
159 & \phm{1}4 & 1.06 & 0.85 & 0.65 & 0.53 & 1.61 & 1.23 \nl
\enddata
\tablecomments{
Columns: (1) MACHO field number, 
(2) number of events from DIA,
(3) total microlensing optical depth from AD2000,
(4) \& (5) {\em correct} asymmetric errors in microlensing optical
depth, 
(6) Gaussian approximation to a Poisson confidence interval based on the
number of events,
(7) \& (8) the ratios of errors: columns (4) or (5) divided by
column (6). 
Similarly to
the optical depth from column (3), the values given in columns 
(4)---(6) are expressed in units of $10^{-6}$.
}
\end{deluxetable}

\begin{deluxetable}{ccccccc}
\tablecaption{Optical Depth in Individual Fields\label{table2}}
\tablewidth{0pt}
\tablehead{
\colhead{Field} & 
\colhead{$N_{DIA}$} &
\colhead{$l\; [^{\circ}]$} &
\colhead{$b\; [^{\circ}]$} &
\colhead{$f_{\rm disk}$} &
\colhead{$\tau_{\rm tot} (\times 10^{6})$} &
\colhead{$\tau_{\rm bar} (\times 10^{6})$}
}
\startdata 
{\bf 101} & 11 & 3.728 & $-3.021$ & 0.118 & $1.72^{+0.61}_{-0.54}$ & $1.95^{+0.69}_{-0.61}$ \nl
104 & 16 & 3.109 & $-3.008$ & 0.106 & $4.18^{+1.65}_{-1.37}$ & $4.68^{+1.84}_{-1.53}$ \nl
{\bf 108} & 16 & 2.304 & $-2.649$ & 0.114 & $2.39^{+0.68}_{-0.61}$ & $2.70^{+0.76}_{-0.69}$ \nl
{\bf 113} & 17 & 1.629 & $-2.781$ & 0.089 & $1.96^{+0.54}_{-0.49}$ & $2.15^{+0.59}_{-0.54}$ \nl
{\bf 118} & 13 & 0.833 & $-3.074$ & 0.100 & $2.64^{+0.97}_{-0.84}$ & $2.93^{+1.08}_{-0.93}$ \nl
{\bf 119} & 12 & 1.065 & $-3.831$ & 0.093 & $2.43^{+0.95}_{-0.81}$ & $2.68^{+1.05}_{-0.89}$ \nl
{\bf 128} & 10 & 2.433 & $-4.029$ & 0.084 & $1.62^{+0.62}_{-0.54}$ & $1.77^{+0.67}_{-0.59}$ \nl
159 & \phm{1}4 & 6.353 & $-4.402$ & 0.093 & $1.06^{+0.85}_{-0.65}$ & $1.17^{+0.94}_{-0.72}$ \nl
\enddata
\tablecomments{
Columns: (1) MACHO field number, 
(2) number of events from DIA,
(3) \& (4) location of the field,
(5) fraction of disk stars along the line-of-sight to $V=23$ from
AD2000 (their factor $p$),
(6) total microlensing optical depth from AD2000 with corrected errors,
(7) microlensing optical depth toward sources in the Galactic bar.\\
The optical depth in column (7) is estimated from the data
given in columns (6) and (5). Boldfaced fields are used in the final 
estimate of the optical depth toward sources residing in the Galactic bar.}
\end{deluxetable}

\clearpage

\begin{table}
\begin{center}
\begin{tabular}{l|lcc}\hline
& Method & Clump giants & DIA \\
& Optical depth & $1.4 \pm 0.3$ & $2.01^{+0.34}_{-0.32}$ \\
& Location $(l,b)$ & (3\hbox{$.\!\!^\circ$}9,
$-3$\hbox{$.\!\!^\circ$}8) & 
(2\hbox{$.\!\!^\circ$}22, $-3$\hbox{$.\!\!^\circ$}18)\\
& Sources & bar & bar+disk \\
Models & Lenses & bar+disk & bar+disk \\
\hline
Binney et al. (1997) && 0.9 (1.2) & 1.0 (1.1) \\
Freudenreich (1998) && 2.0 (2.4) & 1.8 (2.2) \\
Dwek et al.\ (1995) && 1.2 (1.5) & 1.3 (1.6) \\
&&&\\
Bissantz \& Gerhard (2002) && (1.27) & (1.23) \\
\hline
\end{tabular}
\end{center}
\caption{The optical depth in units of $10^{-6}$ for three Galactic models
considered by Evans \& Belokurov (2002) and the non-parametric model
of Bissantz \& Gerhard (2002).
The optical depth values for clump location are taken directly from
Evans \& Belokurov (2002) or Bissantz \& Gerhard (2002), whereas the 
unpublished values at DIA
location were kindly provided by Wyn Evans and Ortwin Gerhard. 
The numbers in parentheses
include the effect of the spiral structure. Note that clump giants
and DIA events probe different populations of microlensing sources.
Observational constraints are given in the top part of the table.
As explained in the main text, the interpretation of this comparison
is difficult.}
\label{table3}
\end{table}

\clearpage

\begin{figure}[htb]
\includegraphics[width=14cm]{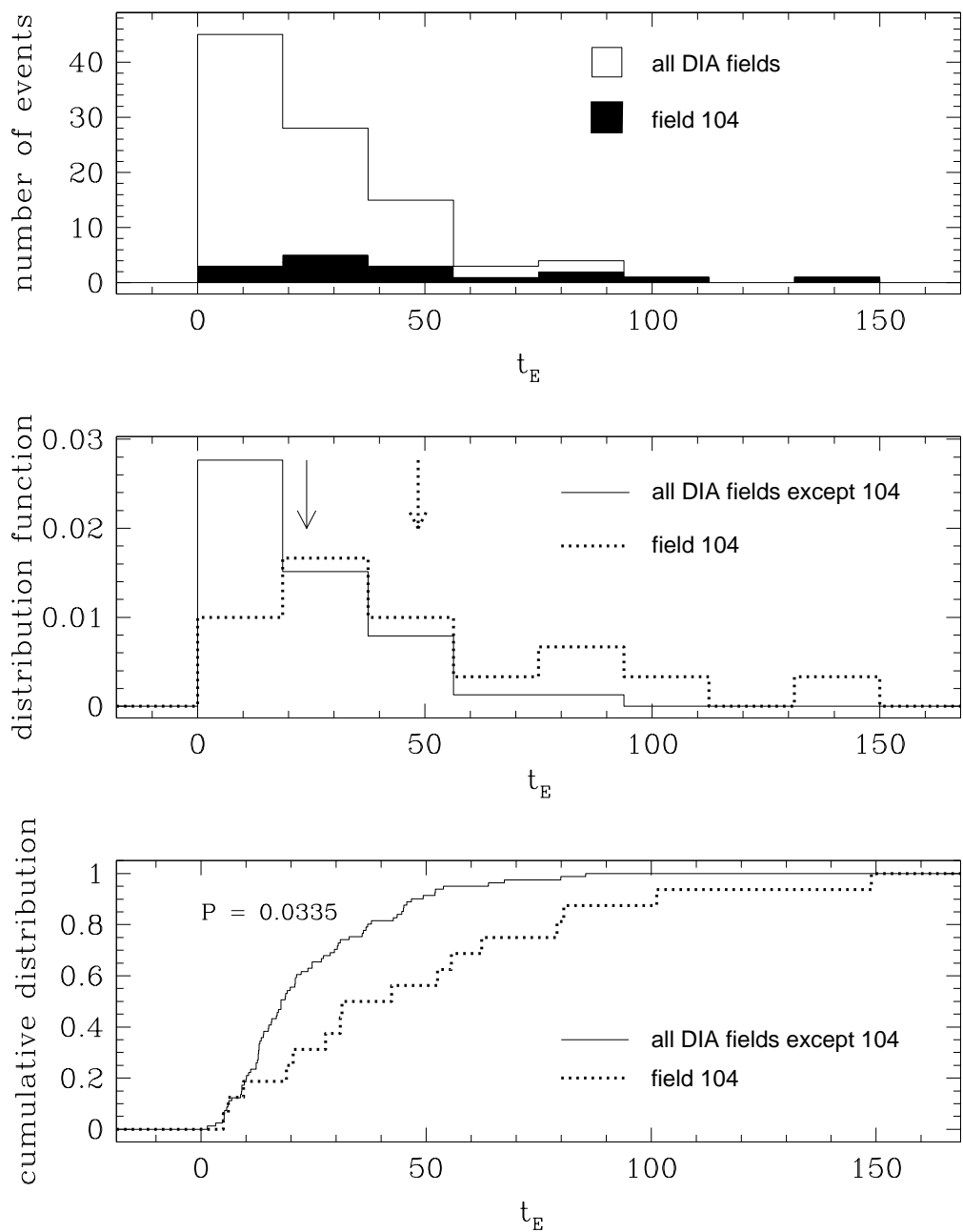}
\caption{The {\em upper panel} shows number of events as a function of event
duration based on all 8 DIA fields (97 events from the sample of 99).
Field 104 clearly dominates the long duration part of this
distribution.
The {\em middle panel} presents comparison of the number distribution function
in field 104 and the remaining 7 fields. The distribution in field
104 is much flatter and more extended. Average durations for both
distributions are marked with vertical arrows.
The {\em lower panel} is a plot of the cumulative distributions of 
event durations
for field 104 and the remaining fields. Kolmogorov-Smirnov test
indicates that the probability that both samples come from the same
parent population is $P = 0.0335$. 
}
\end{figure}

\clearpage

\begin{figure}[htb]
\includegraphics[width=16cm]{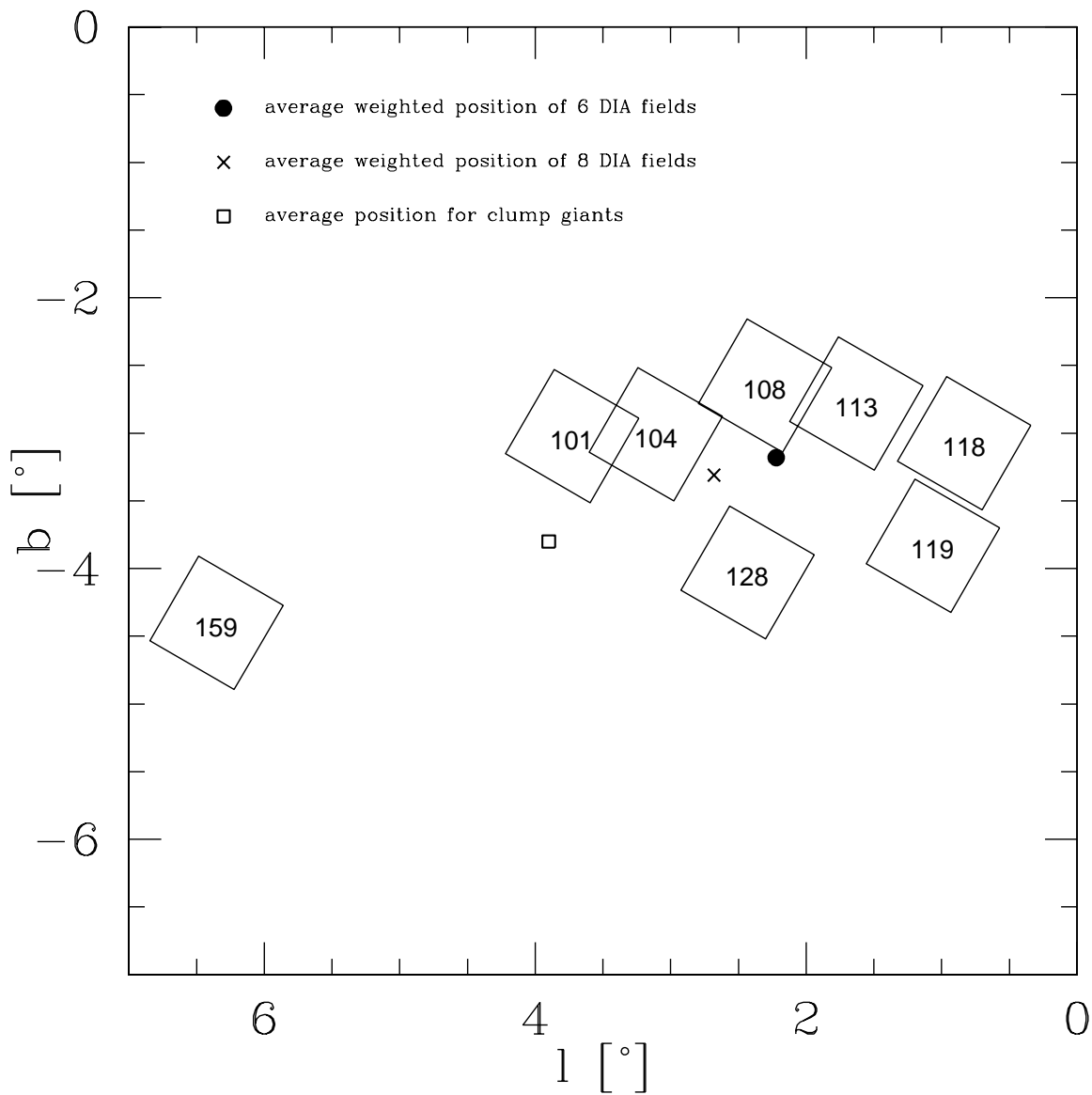}
\caption{
Location of 8 MACHO fields used by AD2000. Field 104
is highly anomalous and field 159 is clearly separated from the
others. The filled circle indicates an average weighted position of 6
fields, at which I give a new optical depth of $\tau_{\rm bar} =
2.23^{+0.38}_{-0.35} \times
10^{-6}$ for $f_{\rm disk} = 0.1$. The average weighted
position of 8 fields is marked with a cross for comparison. The open square
indicates the average position for optical depth of $\tau_{\rm bar} =
1.4 \pm 0.3 \times
10^{-6}$ based on clump giants (Popowski et al.\ 2001).
}
\end{figure}

\clearpage

\end{document}